\documentstyle[subfigure,epsfig,preprint,aps]{revtex}
\newcommand{\bep}{\begin{picture}(200,260)}
\newcommand{\eep}{\end{picture}}

\def\lsim{\raise0.3ex\hbox{$\;<$\kern-0.75em\raise-1.1ex
\hbox{$\sim\;$}}}
\def\gsim{\raise0.3ex\hbox{$\;>$\kern-0.75em\raise-1.1ex
\hbox{$\sim\;$}}}
\newcommand{\vecvar}[1]{\mbox{\boldmath$#1$}}

\begin{document}

\draft

\preprint{}

\title{Two Pion Correlations as a Possible Experimental Probe for
Disoriented Chiral Condensates}

\author{Hideaki Hiro-Oka}

\address{Institute of Physics, Center of Liberal Arts and Sciences, 
Kitasato University \\ 
1-15-1 Sagamihara, Kanagawa 228-8555, Japan} 

\author{Hisakazu Minakata}
\address{Department of Physics, Tokyo Metropolitan University \\
1-1 Minami-Osawa, Hachioji, Tokyo 192-0397, Japan}

\date{Revised in February, 1998}
\preprint{
\parbox{5cm}{
TMUP-HEL-9714\\
hep-ph/9712476\\
}}
\maketitle

\begin{abstract}
We discuss two-pion correlations as a possible experimental
probe into disoriented chiral condensates. In particular, we
point out that the iso-singlet squeezed states of the BCS type 
have peculiar two-particle correlations in the back-to-back and 
the identical momentum configurations which should be detectable 
experimentally. We motivate the examination of the squeezed 
state by showing that such state naturally appears in a final 
stage of nonequilibrium phase transitions via the parametric 
resonance mechanism proposed by Mr\'owczy\'nski and M\"uller.

\end{abstract}

\newpage

The disoriented chiral condensate (DCC)\cite {DCC} is an 
intriguing idea for explaining possible large fluctuations 
of neutral to charge ratio which might be seen in the cosmic 
ray experiments \cite{centauro}. 
A possible picture of DCC is as follows; 
The hot debris formed in high-energy hadronic collisions rapidly 
cools down so that chirally misaligned metastable ``vacuum'' like 
states are formed. They then relax to the QCD vacuum configuration 
by coherently emitting pions, and/or by dynamically producing 
pions via the parametrically amplified resonance mechanism. 

Experimental search for the formation of DCC has been carried 
out by the MiniMax experiment at Fermilab \cite{minimax}. 
Thanks to the well controlled experimental conditions of the 
accelerator experiments, they have a good chance of seeing the 
signature of DCC by observing large neutral to charge fluctuations. 
At least up to this moment, however, a clear signature for 
DCC does not appear to be found. 

The experimental hunting of the signature of DCC so far relies 
on the analysis of global quantities such as multiplicity 
distributions and the correlations between charged and 
neutral particles. Let us consider the following possibility: 
Suppose that DCC forms in hadronic collisions but with a tiny 
probability, for example 1\% or even 0.1\%. 
Then, one may ask the question ``is it possible for such global 
analysis at hand to signal the DCC formation?''. 
The answer is likely to be no. This is the real worry because no 
one ever made attempt to compute, or even estimate, the formation 
probability of DCC in hadronic collisions. 

We propose in this paper that two-pion correlations may serve 
as a good indicator for DCC.\footnote
{The two-particle correlations in the context of DCC has been briefly
discussed in Ref. \cite {muller}, but with somewhat different features 
from those we will obtain below.} 
The two-particle correlation functions 
are experimentally measurable quantities and have been used as a  
sensitive probe for clustering in multiparticle final states 
\cite{cluster}, and for measuring size of the clusters by 
the quantum mechanical interference between identical particles, 
the Hanbury-Brown-Twiss effect \cite{HBT}. 

Two-particle correlations can work as an appropriate measure for 
DCC formation provided that the two-particle correlations in 
the DCC events have different characteristic features from those 
of non-DCC events. We argue that it indeed is the case for broad class 
of models in which the produced multi-pion states can be described 
by the squeezed states. 

It has been suggested by Amado and Kogan \cite{AK} that the quantum 
pion states produced in DCC can be described by the squeezed state. 
The global quantities such as multiplicity distributions and 
correlations are calculated without specifying pion charges in 
Refs. \cite{AK,DH}. 

We explore in this paper a different mechanism which leads to 
the iso-singlet squeezed state with momentum correlation 
of the BCS type. It leads to definite predictions of the two-pion 
correlations with various charge and momentum combinations. 
Although not entirely model-independent, measuring the two-pion 
correlation functions may provide an alternative strategy for 
hunting the formation of DCC. 

Toward the goal we organize this paper in the following two steps: \\
\noindent
(A) We discuss a concrete model proposed by Mr\'owczy\'nski and 
M\"uller \cite{MM} based on the parametric resonance mechanism 
to draw a physical picture behind the multi-pion squeezed state.\\
\noindent
(B) We then extract the characteristic features of two-pion 
correlations implied by the squeezed pion state itself which is 
independent of the dynamical model behind the states. 
A reader who is keenly interested in relatively model-independent 
consequences of the iso-singlet pion squeezed state can skip the 
part (A).\\

Let us start by introducing the model of multipion production 
from DCC to make our discussions concrete. 
We treat the linear $\sigma$ model within the approximation 
discussed by Mr\'owczy\'nski and M\"uller \cite{MM}. 
We define the dynamical variables $\chi$ and $\vec{\pi}$ by expanding 
the $\sigma$ model fields around the minimum of the tilted wine-bottle 
potential as $\sigma = v +\chi$ and $\vec{\pi} = \vec{\pi}$. 
We further expand the dynamical variables around the harmonic 
background of $\chi$ as 
\begin{eqnarray}
\chi(\vecvar{x},t) 
&=& \chi_0 \cos(m_{\sigma}t + \phi) + \hat{\chi}(\vecvar{x},t),\nonumber\\
\vec{\pi}(\vecvar{x},t) 
&=& \hat{\vec\pi}(\vecvar{x},t).\nonumber
\end{eqnarray}
By doing this, we take an intuitive semiclassical picture for 
nonequilibrium phase transition in which the sigma-model 
fields roll down along the $\sigma$ direction and oscillate 
around a minimum of the wine-bottle potential. 

We use a Fourier transformed variable $Q_{\vecvar{k}}$ defined by 
\[
\hat{\chi}(\vecvar{x},t) = 
\displaystyle\int\frac{d^3 k}{(2\pi)^3}
e^{i\vecvar{k}\cdot \vecvar{x}} Q_{\vecvar{k}} (t).
\]
For compactness, we always omit the similar expressions for 
pions whenever they arise in an analogous way of $\sigma$'s.
In spite of the continuum notation above and hereafter we will 
be actually dealing with the theory quantized in a large but 
finite box. We hope that no confusion occurs with our notation. 
The Hamiltonian can be expressed in terms of $Q_{\vecvar{k}}$ 
and its canonical conjugate 
$P_{\vecvar{k}} = \dot{Q}_{-\vecvar{k}}$ as 
\[
H = \displaystyle\int d^3 k \left[
\frac{1}{2} P_{\vecvar{k}} P_{-\vecvar{k}} + 
\frac{1}{2} \Omega_{\vecvar{k}}^{\sigma} (t)^2 
Q_{\vecvar{k}} Q_{-\vecvar{k}} \right] + 
\sum_a \int d^3 k 
\left[
\frac{1}{2}P_{\vecvar{k}}^a P_{-\vecvar{k}}^a + 
\frac{1}{2}\Omega_{\vecvar{k}}^{\pi} (t)^2
Q_{\vecvar{k}}^a Q_{-\vecvar{k}}^a \right],
\]
where 
\begin{eqnarray}
\Omega_{\vecvar{k}}^{\sigma}(t)^2 
&=& \vecvar{k}^2 + m_{\sigma}^2 + 
3 m_{\sigma}^2\left(\frac{\chi_0}{v}\right)\cos(m_{\sigma}t + \phi),
\nonumber\\
\Omega_{\vecvar{k}}^{\pi}(t)^2 
&=& \vecvar{k}^2 + m_{\pi}^2 + 
m_{\sigma}^2\left(\frac{\chi_0}{v}\right)\cos(m_{\sigma}t + \phi).
\nonumber
\end{eqnarray}
The index $a$ runs over 1-3 and we take the adjoint
representation for the pion fields.

We restrict ourselves in this paper to the region of small oscillations, 
$\displaystyle\frac{\chi_0}{v} 
< \left(\frac{m_{\pi}}{m_{\sigma}}\right)^2$, 
so that the time-dependent frequency 
$\Omega_{\vecvar{k}}^{\pi} (t)$ for pions is always real. 
Unless this restriction is made $\Omega_{\vecvar{k}}^{\pi} (t)$ 
becomes imaginary for certain period of time at low momenta 
and it gives rise to another instability in the system. 
We will discuss this problem in our forthcoming publication. Under 
the above restriction the frequency for $\sigma$ automatically 
stays real. 

We follow Shtanov, Traschen and Brandenberger \cite {STB} to obtain 
the quantum description of the multi-pion (and $\sigma$) states 
produced through the parametric resonance mechanism. 
The Hamiltonian can be diagonalized by using the time-dependent 
creation and the annihilation operators
\begin{eqnarray}
a_{\vecvar{k}}(t) &=& 
\displaystyle\frac{1}{\sqrt{2\Omega_{\vecvar{k}}}}
e^{i\int^t dt'\Omega_{\vecvar{k}}(t')} [\Omega_{\vecvar{k}} Q_{\vecvar{k}} 
+ i P_{-\vecvar{k}} ], \nonumber\\
a_{\vecvar{k}}^{\dagger}(t) &=& 
\displaystyle\frac{1}{\sqrt{2\Omega_{\vecvar{k}}}}
e^{-i\int^t dt'\Omega_{\vecvar{k}}(t')} [\Omega_{\vecvar{k}} Q_{-\vecvar{k}} 
- i P_{\vecvar{k}} ], \nonumber
\end{eqnarray}
and the analogous expressions for pions with superscript $a$ as 
\[
H= \int d^3 k (\Omega_{\vecvar{k}}^{\sigma} 
a_{\vecvar{k}}^{\dagger} a_{\vecvar{k}} + 
\sum_a \Omega_{\vecvar{k}}^{\pi} a_{\vecvar{k}}^{a\dagger} a_{\vecvar{k}}^a ).
\]

We now discuss the sigma and the pion sector collectively because 
they are decoupled with each other. 
The Heisenberg equation of motion 
\[
\dot{a}_{\vecvar{k}} (t) = \displaystyle\frac{\partial}{\partial t}
a_{\vecvar{k}}(t) + i [H, a_{\vecvar{k}}]
\]
satisfied by the operators 
$a_{\vecvar{k}} (t)$ and $a_{\vecvar{k}}^{\dagger} (t)$ takes the form 
\begin{equation}
\dot{a}_{\vecvar{k}} (t) = \displaystyle\frac{\dot{\Omega}_{\vecvar{k}}}
{2\Omega_{\vecvar{k}}} e^{2 i \int^t dt'\Omega_{\vecvar{k}}(t')}
a_{-\vecvar{k}}^{\dagger}(t)
\label{Heisenberg}
\end{equation}
and its hermitian conjugate for $a_{\vecvar{k}}^{\dagger}$, where the 
dot implies the time derivative. 
The solution of these equation can be obtained by introducing 
the time-independent operators 
$b_{\vecvar{k}}$ and $b_{\vecvar{k}}^{\dagger}$ as 
\begin{eqnarray}
a_{\vecvar{k}} (t) &=& \alpha_{\vecvar{k}}(t)b_{\vecvar{k}} + 
\beta_{\vecvar{k}}^* (t)b_{-\vecvar{k}}^{\dagger}, \nonumber\\
a_{\vecvar{k}}^{\dagger} (t) &=& \beta_{\vecvar{k}}(t)b_{-\vecvar{k}} + 
\alpha_{\vecvar{k}}^* (t)b_{\vecvar{k}}^{\dagger}. 
\label{Bogoliubov}
\end{eqnarray}
The commutation relation $[b_{\vecvar{k}}, b_{\vecvar{k'}}^{\dagger}]=
\delta_{\vecvar{k},\vecvar{k'}}$ implies that 
\[
|\alpha_{\vecvar{k}}|^2 - |\beta_{\vecvar{k}}|^2 = 1.
\]
It is nothing but the Bogoliubov transformation but one must notice 
that $b_{\vecvar{k}}$ and $b_{\vecvar{k}}^{\dagger}$ do {\it not} 
diagonalize the Hamiltonian. 
They are introduced to solve the Heisenberg equation 
(\ref{Heisenberg}) but not to diagonalize the Hamiltonian. 
The $a_{\vecvar{k}}$ and $a^{\dagger}_{\vecvar{k}}$ in (\ref{Bogoliubov}) 
solve the equation (\ref{Heisenberg}) if 
$\alpha_{\vecvar{k}}$ and $\beta_{\vecvar{k}}$ obey the equations 
\begin{eqnarray}
\dot{\alpha}_{\vecvar{k}}(t') 
&=& \displaystyle
\frac{\dot\Omega_{\vecvar{k}}}{2\Omega_{\vecvar{k}}}
e^{2i \int^t dt'\Omega_{\vecvar{k}}(t')} \beta_{-\vecvar{k}},
\nonumber\\
\dot{\beta}_{\vecvar{k}}(t) 
&=& \displaystyle\frac{\dot\Omega_{\vecvar{k}}}
{2\Omega_{\vecvar{k}}}e^{-2i \int^t dt'\Omega_{\vecvar{k}}(t')} 
\alpha_{-\vecvar{k}}\nonumber
\end{eqnarray}

We define the vacua $|0 \rangle$ and $|0(t) \rangle$ as 
\begin{eqnarray}
b_{\vecvar{k}}|0 \rangle &=& 0,\nonumber\\
a_{\vecvar{k}}(t) |0(t) \rangle &=& 0,\nonumber
\end{eqnarray}
and require that $|0(t=0) \rangle = |0\rangle$. 
It leads to the initial conditions 
\begin{equation}
\alpha_{\vecvar{k}}(0) =1, \; \beta_{\vecvar{k}}(0) =0.
\label{boundary}
\end{equation}
A physical quantum of momentum $\vecvar{k}$ is created (annihilated) 
by $a_{\vecvar{k}}^{\dagger}(t)$ ($a_{\vecvar{k}}(t)$) out of 
the vacuum $|0(t) \rangle$ because they are the variables that  
diagonalize the Hamiltonian at every time $t$. 
On the other hand, the quantum sigma state in DCC 
may be identified by $|0 \rangle$. 

Because of the Bogoliubov transformation (\ref{Bogoliubov}) 
the $\sigma$ state $|0 \rangle$ in DCC 
can be expressed by physical $\sigma$ quanta as
\begin{equation}
|0 \rangle = \prod_{\vecvar{k}} \displaystyle\frac{1}
{\sqrt{|\alpha_{\vecvar{k}}(t)|}}\mbox{exp} \left[
\frac{\beta^*_{-\vecvar{k}}}{2\alpha^*_{\vecvar{k}}} a^{\dagger}_{\vecvar{k}}
(t) a^{\dagger}_{-\vecvar{k}}(t)\right] |0(t) \rangle.
\label{squeezed}
\end{equation}
It may be appropriate to denote it as the squeezed state of the 
BCS type, as we do in this paper, because of the pairing between
$\vecvar{k}$ and $-\vecvar{k}$ modes.
The states of the generic type of (\ref{squeezed}) are the 
squeezed states that is widely discussed in quantum optics 
\cite{optics}. We also note that the particle production via 
the parametric resonance has been extensively discussed in the 
context of reheating in the inflationary universe 
\cite{STB,linde,boyan,yoshim}. 

We write the solution of the Heisenberg equation 
\begin{equation}
\ddot{Q}_{\vecvar{k}}(t) + \Omega^2_{\vecvar{k}}(t)Q_{\vecvar{k}}(t) =0
\label{Mathieu}
\end{equation}
as
\[
Q_{\vecvar{k}}(t) = Q^{(-)}_{\vecvar{k}}(t)b_{\vecvar{k}} + 
Q^{(+)}_{\vecvar{k}}(+) b_{-\vecvar{k}}^{\dagger}.
\]
Where $Q^{(-)}_{\vecvar{k}}$ and $Q^{(+)}_{\vecvar{k}}$ are the solutions 
of the classical Mathieu equation (\ref{Mathieu}) \cite {Mathieu}. 
They satisfy $Q^{*(-)}_{\vecvar{k}} = Q^{(+)}_{-\vecvar{k}}$. 
It is easy to show that $\alpha_{\vecvar{k}}$ and $\beta_{\vecvar{k}}$ 
can be expressed as 
\begin{eqnarray}
\alpha_{\vecvar{k}}(t) &=& 
\displaystyle\frac{1}{\sqrt{2\Omega_{\vecvar{k}}}}
e^{i \int^t dt' \Omega_{\vecvar{k}}} \left[
\Omega_{\vecvar{k}} Q^{(-)}_{\vecvar{k}} + 
i\dot Q^{(-)}_{\vecvar{k}} \right],\nonumber\\
\beta_{-\vecvar{k}}(t) &=& 
\displaystyle\frac{1}{\sqrt{2\Omega_{\vecvar{k}}}}
e^{-i \int^t dt' \Omega_{\vecvar{k}}} \left[
\Omega_{\vecvar{k}} Q^{(-)}_{\vecvar{k}} - 
i\dot Q^{(-)}_{\vecvar{k}} \right].\nonumber
\end{eqnarray}
By including the pion degrees of freedom the quantum state of DCC in 
the Mr\'owczy\'nski-M\"uller model can be written as
\begin{equation}
|0 \rangle^{\sigma} \otimes |0 \rangle^{\vec{\pi}} 
\equiv |\psi \rangle.
\label{ketpsi}
\end{equation}
One of the most important feature of the state (\ref{ketpsi}) is that 
it is the iso-singlet state. It comes from the fact that the 
frequency $\Omega_{\vecvar{k}}^{\pi}$ is isospin singlet.

Once the state is specified it is easy to compute the expectation 
values of the number densities of pion and $\sigma$ quanta. 
We use the collective notation 
\[
\langle n \rangle^{\alpha}_{\vecvar{k}} \equiv
\langle \psi| {a_{\vecvar{k}}^{\alpha}}^{\dagger}(t)a_{\vecvar{k}}^{\alpha}(t)
|\psi \rangle,
\]
where $\alpha$ runs $\sigma$ and $\pi^a (a = +,-,0)$. Here, 
we use the representation of pion fields with definite charge states. 
Then, 
\[
\langle n \rangle^{\alpha}_{\vecvar{k}} = 
|\beta_{\vecvar{k}}^{\alpha}|^2.
\]
We will comment on the question of how we should interpret the 
production of $\sigma$ quanta at the end of this paper. 

We give in Fig. 1 the single pion momentum distributions calculated 
by the parametric resonance mechanism. They are identical for 
$\pi^+$, $\pi^-$ and $\pi^0$ because $|\psi \rangle$ is isospin singlet. 
The $\sigma$ model parameters are taken as follows: 
$m_{\pi} = 140$ MeV, $m_{\sigma} = 600$ MeV and 
$\displaystyle\frac{\chi_0}{v} = \frac {1}{20}$. 
The function $\beta_{\vecvar{k}}$ is computed by solving the 
Mathieu equation subject to the boundary condition (\ref{boundary}). 
We observe a sharp peak at around $k = 270$ MeV which is attributable 
to the first instability band of the Mathieu equation.
An another peak due to the second resonance band which should 
exist at around $k = 580$ MeV is too sharp to be visible in a 
finite-size binning.
If we take the larger value of $\displaystyle\frac{\chi_0}{v}$ 
the peak widths become wider. If these spikes are not masked by 
the non-DCC backgrounds then it provides a sensitive experimental 
signature for the parametric resonance mechanism.


Now we move on to the part (B) to address the observable consequences
of the squeezed state of the BCS type (\ref{ketpsi}). 
We are interested in two-pion correlations. 
To this goal we first compute the two-particle momentum 
distributions of pions. They are defined as 
\[
\langle \pi^a_{\vecvar{k}}, \pi^b_{\vecvar{k'}} \rangle \equiv 
\langle \psi| a^{a\dagger}_{\vecvar{k}}(t) a^a_{\vecvar{k'}}(t) 
a^{b\dagger}_{\vecvar{k'}}(t) a^b_{\vecvar{k'}}(t) | \psi\rangle.
\]
Because of the factorized form of the state $|0\rangle^{\vec\pi}$ 
as in (\ref{squeezed}) we expect that the nontrivial 
(= not manifestly factorizable) two-pion distributions arise 
only in the sector of the identical and the back-to-back momentum 
configurations. 
In the computation of these quantities we further recognize that 
the two-pion distributions at zero-momentum can {\it not} be obtained 
by taking the smooth limit $\vecvar{k}\rightarrow 0$ of the 
expressions of either identical or back-to-back momentum
configurations.
It is because $a_{\vecvar{k}=0}$ and $a^{\dagger}_{\vecvar{k}=0}$ 
do not commute, whereas $a_{\vecvar{k}}$ and $a^{\dagger}_{-\vecvar{k}}$ 
do commute for $\vecvar{k} \neq 0$. Therefore, we have to compute 
three types of the two-pion distributions separately. 
We only quote the result leaving the details (which are not 
difficult to work out) to our forthcoming paper \cite{HM}. \\
\noindent
(a) identical momentum distribution:
\begin{eqnarray}
\langle \pi^{+}_{\vecvar{k}},\pi^{+}_{\vecvar{k}} \rangle &=&
\langle \pi^0_{\vecvar{k}},\pi^0_{\vecvar{k}} \rangle =
\langle n \rangle_{\vecvar{k}} 
\left(2\langle n \rangle_{\vecvar{k}} +1 \right),
\nonumber\\
\langle \pi^{+}_{\vecvar{k}},\pi^{-}_{\vecvar{k}} \rangle &=&
\langle n \rangle^2_{\vecvar{k}}.\nonumber
\end{eqnarray}

\noindent
(b) back-to-back momentum distribution: 
\begin{eqnarray}
\langle \pi^{+}_{\vecvar{k}}, \pi^{-}_{-\vecvar{k}} \rangle &=& 
\langle \pi^0_{\vecvar{k}}, \pi^{0}_{-\vecvar{k}} \rangle =
\langle n \rangle_{\vecvar{k}}
\left(2\langle n \rangle_{\vecvar{k}} +1 \right),
\nonumber\\
\langle \pi^{+}_{\vecvar{k}}, \pi^{+}_{-\vecvar{k}} \rangle &=& 
\langle n \rangle^2_{\vecvar{k}}.\nonumber
\end{eqnarray}

\noindent
(c) zero-momentum distribution: 
\begin{eqnarray}
\langle \pi^{+}_{\vecvar{k}=0}, \pi^{+}_{\vecvar{k}=0} \rangle &=& 
\langle \pi^{+}_{\vecvar{k}=0}, \pi^{-}_{\vecvar{k}=0} \rangle =
\langle n \rangle_{\vecvar{k}=0} 
\left(2\langle n \rangle_{\vecvar{k}=0}+1 \right),
\nonumber\\
\langle \pi^{0}_{\vecvar{k}=0}, \pi^{0}_{\vecvar{k}=0} \rangle &=& 
\langle n \rangle_{\vecvar{k}=0} 
\left(3\langle n \rangle_{\vecvar{k}=0} +2 \right).
\nonumber
\end{eqnarray}

The two-pion correlation function is defined by 
\[
C(\pi^a_{\vecvar{k}}, \pi^b_{\vecvar{k'}}) \equiv 
\langle \pi^a_{\vecvar{k}}, \pi^b_{\vecvar{k'}} \rangle -
\delta_{ab}\delta_{\vecvar{k},\vecvar{k'}} \langle\pi^a_{\vecvar{k}} \rangle - 
\langle \pi^a_{\vecvar{k}} \rangle \langle \pi^b_{\vecvar{k'}}\rangle.
\]
Then, it is straightforward to obtain the following results. 

\noindent
(a) identical momentum correlations: 
\begin{eqnarray}
C(\pi^{+}_{\vecvar{k}},\pi^{+}_{\vecvar{k}}) &=& 
C(\pi^0_{\vecvar{k}}, \pi^0_{\vecvar{k}}) = 
\langle n \rangle^2_{\vecvar{k}},\nonumber\\
C(\pi^{+}_{\vecvar{k}}, \pi^{-}_{\vecvar{k}}) &=& 0.\nonumber
\end{eqnarray}

\noindent
(b) back-to-back momentum correlations:
\begin{eqnarray}
C(\pi^{+}_{\vecvar{k}}, \pi^{-}_{-\vecvar{k}}) &=& 
C(\pi^0_{\vecvar{k}}, \pi^0_{-\vecvar{k}}) = 
\langle n_{\vecvar{k}} \rangle
\left(\langle n \rangle_{\vecvar{k}} +1 \right),\nonumber\\
C(\pi^{+}_{\vecvar{k}}, \pi^{+}_{-\vecvar{k}}) &=& 0.\nonumber
\end{eqnarray}

\noindent
(c) zero-momentum correlations: 
\begin{eqnarray}
C(\pi^{+}_{\vecvar{k}=0}, \pi^{+}_{\vecvar{k}=0}) &=& 
\langle n \rangle^2_{\vecvar{k}=0}, \nonumber\\
C(\pi^{+}_{\vecvar{k}=0}, \pi^{-}_{\vecvar{k}=0}) &=& 
\langle n \rangle_{\vecvar{k}=0}
\left(\langle n \rangle_{\vecvar{k}=0} +1 \right), \nonumber\\
C(\pi^{0}_{\vecvar{k}=0}, \pi^{0}_{\vecvar{k}=0}) &=& 
\langle n \rangle_{\vecvar{k}=0}
\left(2\langle n \rangle_{\vecvar{k}=0} +1 \right). \nonumber
\end{eqnarray}
As will be discussed in Ref.\cite{HM} the features of the 
correlation functions in (c) can be understood partly as a 
consequence of the isospin singlet nature of (\ref {ketpsi}). 

The distinctive features of the two-pion correlations include: 

\noindent
(i) The back-to-back correlations are stronger than the same-$\vecvar{k}$ 
correlations, as it is natural for the pairing of $\vecvar{k}$ and 
$-\vecvar{k}$ modes in the BCS state (\ref{squeezed}). 

\noindent
(ii) The results of the back-to-back correlations represent solely 
the dynamical character of the mechanism. In particular the sharp peaking 
of the $\pi^{+}$ and $\pi^{-}$ correlation should give an unmistakable 
signature of the squeezed states formed through dynamical pion production 
by parametric resonance. On the other hand, the features of the identical 
momentum correlation may be understood as partly due to the identical 
particle interference. 

\noindent
(iii) The zero-momentum correlations are most dramatic. 
We may argue that they are contributed by the dynamical nature of 
pion correlations due to parametric resonance as well as by the 
identical particle interference; 
Only the latter does not explain $C(\pi^{+}, \pi^{-})$ and the 
fact that $C(\pi^0, \pi^0) > 2C(\pi^{+}, \pi^{+})$. 

To summarize, we have discussed the possibility that the two-pion 
correlations of various definite charge/momentum combinations can be 
used as a new tool for hunting DCC. We also analyzed quantum aspects 
of the Mr\'owczy\'nski-M\"uller model in which pions are produced 
via the parametric resonance mechanism in the later stage of the 
nonequilibrium chiral phase transition. 

Finally, some remarks are in order:

\noindent
(1) If the squeezed state is of non-BCS type with paring of identical 
momentum modes as discussed in \cite{AK}, the prediction to the 
two-pion correlations with identical momentum modes follows that 
of the zero-momentum modes, (c) in our treatment.  

\noindent
(2) In this paper we assumed that the classical background fields 
which rolls down along the $\sigma$ direction toward the minimum of 
the wine-bottle potential. This contrasts with the conventional 
picture of DCC which emphasizes that the rolling-down motion is 
nearly isotropic in isospin space. 
The effect of approximate isotropy of the rolling down motion 
will be discussed in ref. \cite{HM}. Our preliminary result seems 
to indicate that qualitative features of two-pion correlations
remain unaffected. 

\noindent
(3) The $\sigma$ degrees of freedom in the linear $\sigma$ model 
may represent, at least partly, collective multi-pion excitations. 
It also affects the quantitative features of two-pion correlations.
However, we do not know how to implement this effect into our 
calculation. Rather, we prefer to restrict our discussions into 
the qualitative features of two-pion correlations in this paper.

\begin{figure}[t]
\begin{center}
\subfigure[
\baselineskip=15pt
The single pion momentum distribution as a function of the scaled 
time $z= \frac {1}{2}m_\sigma t$. 
The parameters are taken as $m_\pi$=140MeV, $m_\sigma$=600MeV, 
and $\chi_0/v=0.05$. 
]
{\epsfig{file=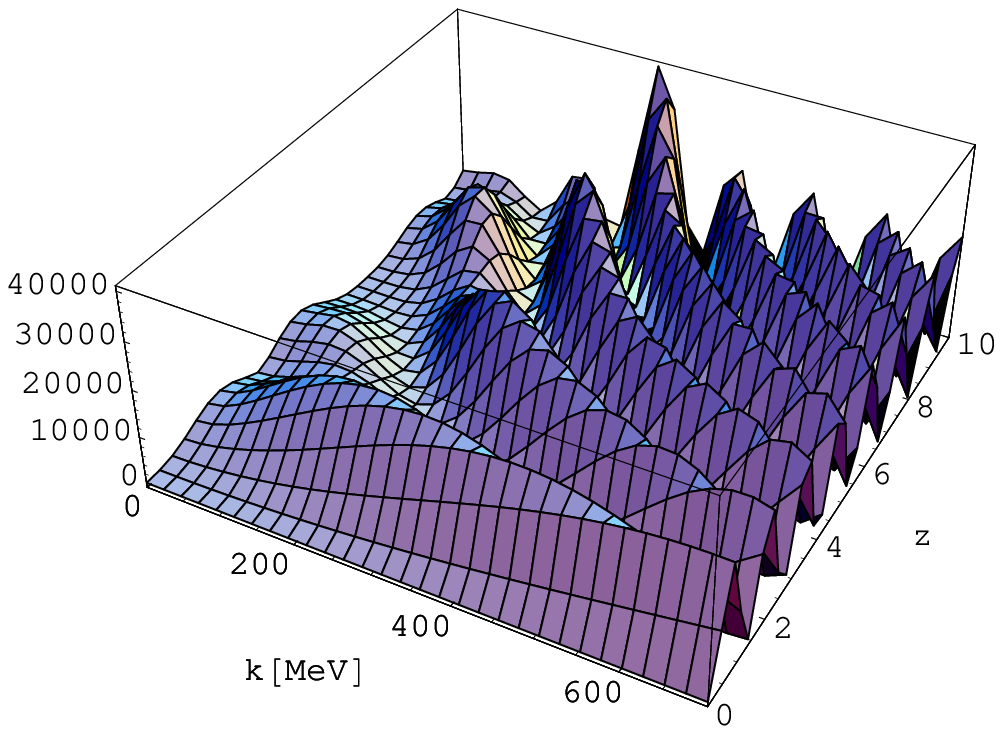,width=13.5cm}}
\subfigure[
\baselineskip=15pt
The single pion momentum distribution integrated over 
time $z$ from 0 to 20, which corresponds to the period of time 
from $t=0$ to 13 fm.
]
{\epsfig{file=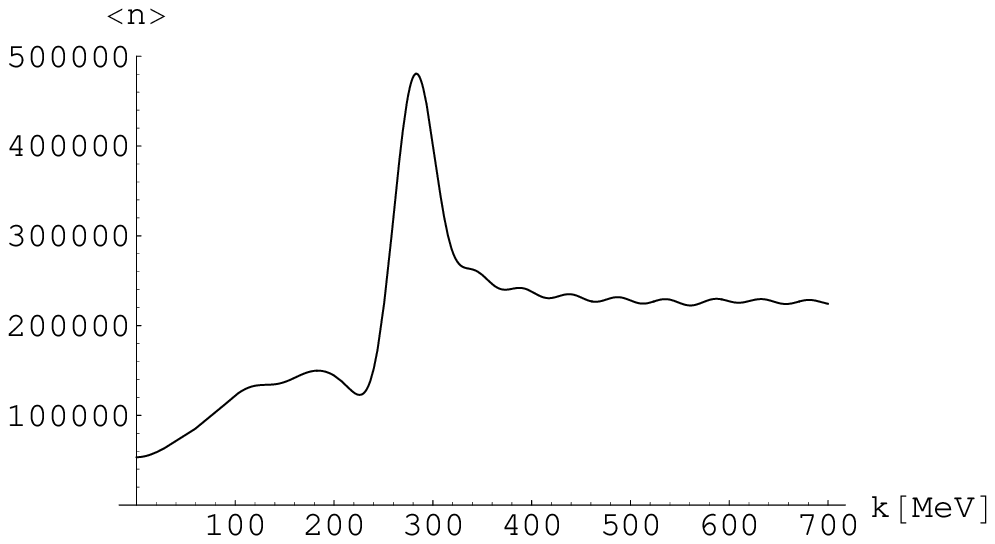,width=13.5cm}}
\end{center}
\end{figure}


\begin{references}

\bibitem {DCC}
A. A. Anselm, Phys. Lett. {\bf B217} (1989) 169;
A. A. Anselm and M. G. Ryskin, Phys. Lett. {\bf B266} (1991) 482;
J.~D.~Bjorken, Acta Phys. Pol. {\bf B23} (1992) 561;
J.~D.~Bjorken, K. L. Kowalski, and C. C. Taylor, Talk at 7th 
Rencontres de Physique de la Valee d'Aoste, 
{\it La Thuile Rencontres 1993}, page 507. 
J.-P. Blaizot and A. Krzywicki, Phys. Rev. {\bf D46} (1992) 246;
Phys. Rev. {\bf D50} (1994) 442;
K.~Rajagopal and F.~Wilczek, Nucl. Phys. {\bf B404} (1993) 577;
S.~Gavin, A.~Gocksch, and R.~D.~Pisarski, Phys. Rev. Lett. {\bf 72} 
(1994) 2143;
M.~Asakawa, Z.~Huang, and X.-N.~Wang, Phys. Rev. Lett.{\bf 74} (1995) 3126;
K.~Rajagopal, Talk at 25th International Workshop on Gross Properties of 
Nuclei and Nuclear Excitation: QCD Phase Transitions, January 13-18, 1997 
Hirschegg, Austria, hep-ph/9703258.

\bibitem {centauro}
L. T. Baradzei et al., Nucl. Phys. {\bf B370} (1992) 365.  
J. Lord and J. Iwai, paper submitted to International Conference 
on High Energy Physics, Dallas, Texas, 1992. 

\bibitem {minimax}
T. Brooks et al. (MiniMax Collaboration), hep-ph/9609375. 
J. Street, Talk at Argonne Workshop on Hadron Systems at High Density 
and/or High Temperature, August 7, 1997.  

\bibitem {cluster}
R. C. Hwa, Summary Talk at 7th International Workshop on Multiparticle 
Production; Correlations and Fluctuations, June 30-July 6, Nijmegen, 
Netherlands, in {\it Multiparticle Physics 1996}, page 377. 

\bibitem {HBT}
R. Hanbury-Brown and R. Q. Twiss, Phil. Mag. {\bf 45} (1954) 633; 
G. Goldhaber, S. Goldhaber, W. Lee, and A. Pais, Phys. Rev. {\bf 120} 
(1960) 300;
E. V. Shuryak, Phys. Lett. {\bf B44} (1973) 387;
Sov. J. Nucl. Phys. {\bf 18} (1974) 667; 
G. I Kopylov and M. I. Podgoretsky, Sov. J. Nucl. Phys. 
{\bf 14} (1971) 1084; {\bf 18} (1973) 656;
G. I Kopylov, Phys. Lett. {\bf B50} (1974) 472.  

\bibitem {muller}
C. Greiner, C. Gong, and B. M\"uller, Phys. Lett. {\bf B316} (1993) 226.

\bibitem {AK}
I. I. Kogan, JETP Lett. {\bf 59} (1994) 312; 
R. D. Amado and I. I. Kogan, Phys. Rev. {\bf D51} (1995) 190.

\bibitem {DH}
I. M. Dremin and R. C. Hwa, Phys. Rev. {\bf D53} (1996) 1216.

\bibitem {MM}
S. Mr\'owczy\'nski and B. M\"uller, Phys. Lett. {\bf B363} (1995) 1.

\bibitem {STB}
Y. Shtanov, J. Traschen and R. Brandenberger, 
Phys. Rev. {\bf D51} (1995) 5438.

\bibitem {optics}
H. P. Yuen, Phys. Rev. {\bf A13} (1976) 2226;
D. F. Walls, Nature {\bf 306} (1983) 141;
D. F. Smirnov and A. S. Troshin, Sov. Phys. Usp. {\bf 30} (1987) 851. 
 
\bibitem {linde}
L. Kofman, A. Linde, and A. A. Starobinsky, Phys. Rev. Lett. 
{\bf 73} (1994) 3195; Phys. Rev. {\bf D56} (1997) 3258.

\bibitem {boyan}
D. Boyanovsky, H. J. de Vega, R. Holman, D.- S. Lee, and A. Singh, 
Phys. Rev. {\bf D51} (1995) 4419;
D. Boyanovsky, H. J. de Vega, R. Holman, and J. F. J. Salgado, 
Phys. Rev. {\bf D54} (1996) 7570.

\bibitem {yoshim}
M. Yoshimura, Prog. Theor. Phys. {\bf 94} (1995) 873.
H. Fujisaki et al., Phys. Rev. {\bf D53} (1996) 6805; 
{\bf D54} (1996) 2494.


\bibitem {Mathieu}
M. Abramowitz and I. A. Stegun, {\it Handbook of Mathematical Functions}
(Dover Publication, New York, 1965).

\bibitem {HM}
H. Hiro-Oka and H. Minakata, in preparation. 

\end{references}
\end{document}